\documentstyle[psfig]{l-aa}
\input epsf
\begin{document}
\thesaurus{08                
	      (02.19.2;      
	       11.08.1;      
	       11.09.1 M82)} 
\title{{\em Letter to the Editor}\\
Observation of Extended Red Emission (ERE) in the halo of M82
\thanks{Based on observations made at the
Observatoire de Haute Provence du CNRS}}
\subtitle{ }
\author{J.-M. Perrin, S. Darbon and J.-P. Sivan}
\offprints{J.-P. Sivan}
\institute{Observatoire de Haute Provence du CNRS, 04870 Saint Michel
l'Observatoire, France}
\date{Received xxx; accepted yyy  }
\maketitle
\begin{abstract}
Low-dispersion spectra have been obtained from 4000 to 9600 \AA~along the axis
and across the dusty halo of the galaxy M82.
The spectra of the halo are explained in terms of scattering of the galactic
light by dust grains:
when divided by the spectrum of the galaxy, they reveal a broad emission band,
similar to, although wider than the bump of Extended Red Emission (ERE)
commonly  observed in galactic reflection nebulae.
The feature is particularly well marked in the filamentary parts of the halo.

This is the first detection of ERE outside of our Galaxy.
\end{abstract}
\section{Introduction}
The so-called Extended Red Emission (ERE) has been observed in the spectrum of
various galactic objects:
reflection nebulae (Witt \& Boronson 1990),
planetary nebulae (Furton \& Witt 1992),
H{\sc ii} regions (Sivan \& Perrin 1993),
the dark nebula L1780 (Mattila 1979, Chlewicki \& Laureijs 1987)
and high latitude cirrus (Guhathakurta \& Tyson 1989).
It consists of a broad emission
bump about 1000 \AA~ wide (FWHM), centered between 6500 and 7500 \AA. The bump
has been explained either by fluorescence from isolated molecules such as
Polycyclic Aromatic Hydrocarbons (PAH) (d'Hendecourt et al. 1986, L\'eger et
al. 1988) or by fluorescence from solid state materials such as
filmy Quenched Carbonaceous Composites (f-QCC) (Sakata et al. 1992)
and Hydrogenated Amorphous Carbon (HAC) (Watanabe et al. 1982,
Furton \& Witt 1993).
Although laboratory experiments  on solid materials
presently provide for the observed ERE
a better fit  than experiments on isolated molecules, no interpretation can be
unambiguously established. Further laboratory studies as well as further
observations of dusty objects are needed for a better understanding of the
physico-chemical processes  responsible for the ERE.
In particular, the question about the universality of the ERE may arise.
The galaxy M82 is a good candidate to attempt to answer
this question because its halo might be considered as a giant
reflection nebula (Schmidt et al. 1976) illuminated by the galaxy itself,
which is seen approximately edge-on.
It is thus easy to compare the spectrum of the galaxy
to that of the light emitted by the halo. We have done such a study:
this Letter describes  our observations, and compare the results to
those of analogous studies on galactic objects.
\begin{figure}[h]
\epsfxsize=9.0cm \epsfysize=6.8cm \epsfbox{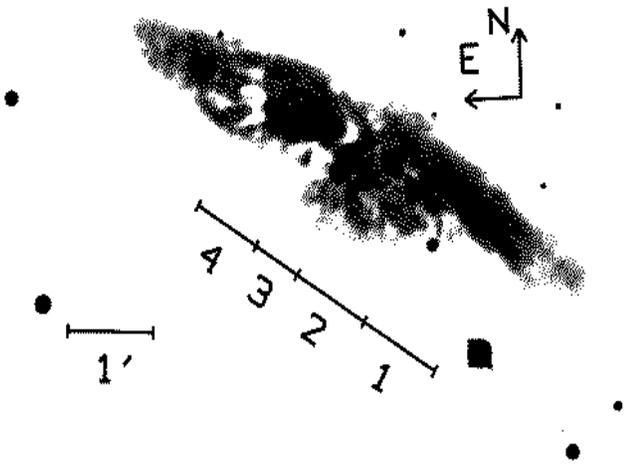}
\caption[]{The four regions observed spectrophotometrically
in the halo of the galaxy M82, drawn  on a photograph obtained
with the 193cm telescope of the Observatoire de Haute Provence}
\label{photo}
\end{figure}
\section{Observations and data reduction}
Spectra of M82 were obtained in February 1995 using the Carelec long-slit
spectrograph (Lema\^{\i}tre et al. 1990)
equiped with a thinned back-illuminated Tektronix CCD, and mounted
at the Cassegrain focus of the 193 cm
telescope of the Observatoire de Haute Provence.
Observations and data reduction were conducted following a procedure similar
to that  previously used for the study of H{\sc ii} regions
(Perrin \& Sivan 1992, Sivan \& Perrin 1993).
The spectral domain 4000 --
9600 \AA~was covered with a dispersion of 277 \AA~mm$^{-1}$. The slit width was
$5\arcsec$.2, wich corresponds to a resolution of 32 \AA.

Two parallel slit positions were observed, the first one along the
axis of the galaxy and the second one across the halo, $100\arcsec$
south-east  (Fig. \ref{photo}).
In the latter case, we made a number of short exposures
alternatively on the halo and on the surrounding sky background. This
procedure permits a fine monitoring of the continuum spectrum of the night-sky
background. After co-adding
the short exposure frames, offset subtracting and flat fielding, the
halo spectrum was spatially divided into four adjacent bins
 to separate  the filaments present in the halo (regions 2, 3 and 4)
from purely diffuse areas (region 1), as shown in Fig. \ref{photo}.
One-dimensional spectra were then extracted for each region.
Sky subtraction was done by removing from halo spectra the corresponding
sky background spectra extracted by binning the same pixels along the
spectrograph slit (thus avoiding any instrumental effect).
The same procedure was applied for the spectrum of the galaxy.
\begin{figure}[h]
\epsfbox{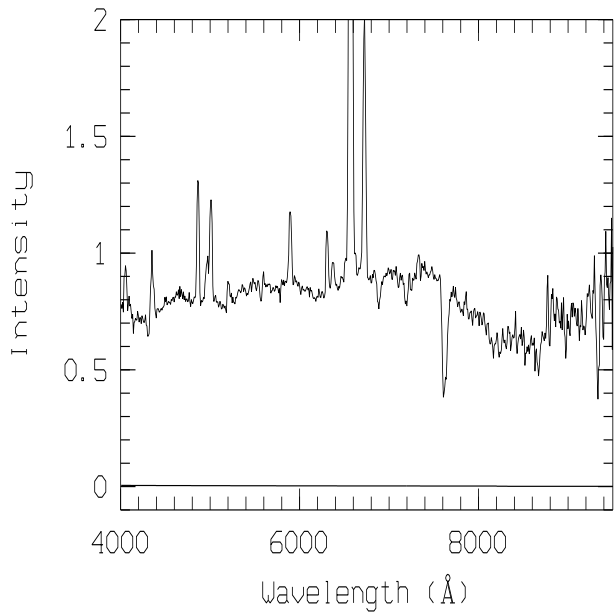}
\caption[]{ The spectrum of region 2 (Fig. \ref{photo}).
The total integration time is of  100 minutes.
Note the emission lines which are scattered lines from the galaxy and,
close to zero level, the atomic continuum for hydrogen, calculated
as explained in text. Ordinates are in units
of $10^{-16}$ erg cm$^{-2}$ s$^{-1}$ \AA$^{-1}$ arcsec $^{-2}$}
\label{brut}
\end{figure}
\section{Results and Discussion}
Figure \ref{brut} displays the spectrum of region 2 which corresponds to
the most intense filament of the halo of M82.
A strong continuum is observed with some superimposed emission lines
such as H$\alpha$ and H$\beta$. According to
Visvanathan \& Sandage (1972) and Scarrot et al. (1991), the H$\alpha$ line
is polarized and is at least partially
due to scattering by grains of the line emitted by the nucleus of the galaxy.
As the polarization is rather high, the atomic component in the observed
continuum should be negligeable. This is confirmed
by theoretical calculations we have conducted for
a pure hydrogen region and for a wide range of
electronic temperatures ($5000  \leq T_e \leq 20000$K)
 and electronic densities ($100 \leq n_e \leq 10000$ cm$^{-3}$):
the atomic continuum, when  normalized to the observed
H$\beta$ intensity, is found to be at the most four
magnitudes lower than the observed continuum (Fig. \ref{brut}).
So, it is clear that the observed continuous spectrum
of the halo of M82 should be regarded as essentially
arising from scattering by grains of the light from the galaxy,
a result in good agreement with that of Schmidt et al. (1976).

\begin{figure}[h]
\epsfxsize=18.0cm \epsfysize=16.0cm
\epsfbox{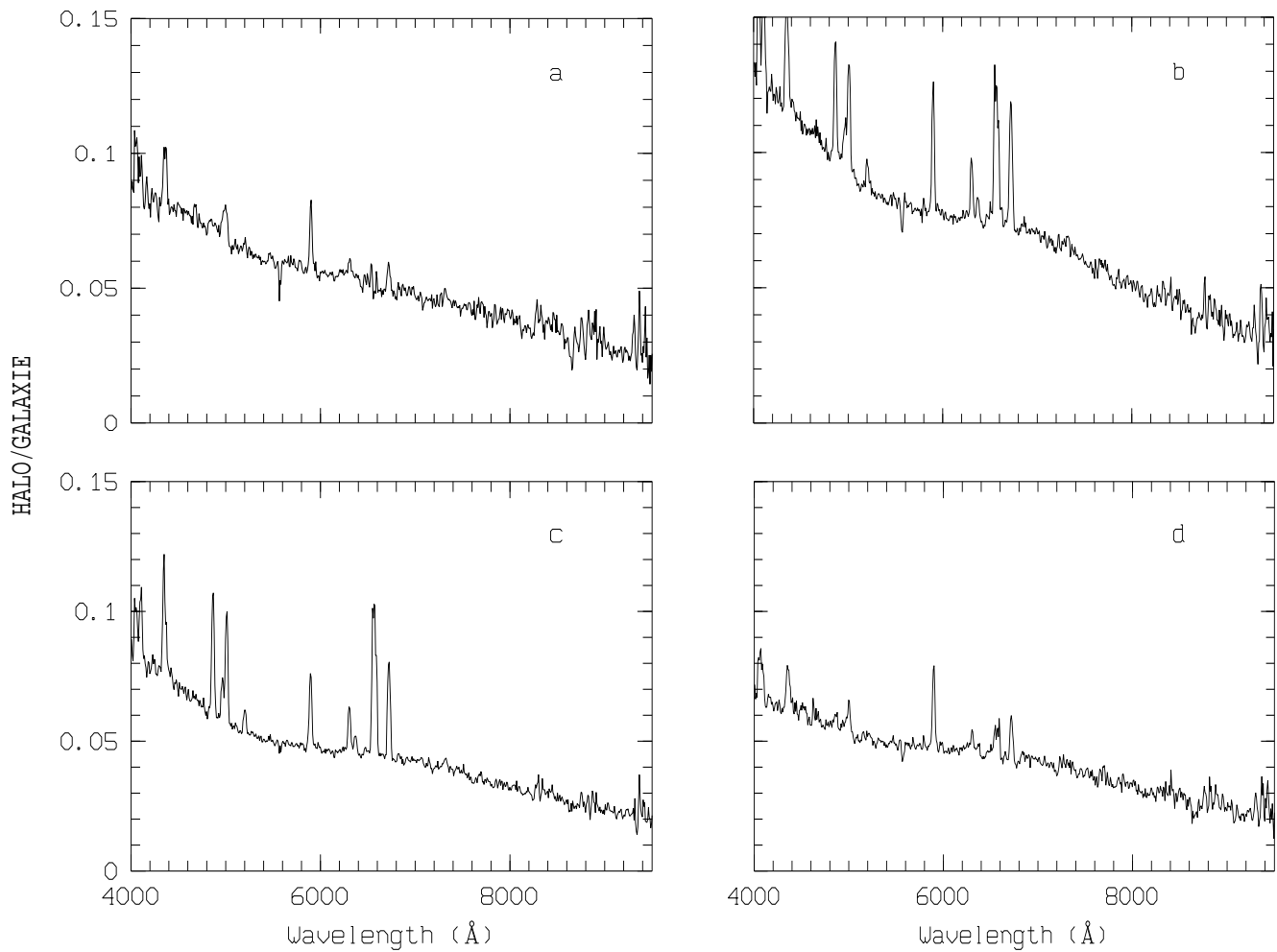}
\caption[]{Ratio of the spectrum of the halo of M82 to  that of the
galaxy, for regions 1 ({\bf a}) and 3 ({\bf c})
(Fig. \ref{photo})}
\label{divis}
\end{figure}

Assuming the same interstellar reddening for the galaxy and its halo,
we have divided the   spectra of the four regions considered in the halo
by the spectrum of the galaxy. The resulting "spectra"  are
 shown in Fig. \ref{divis}: they appear to be typical of dust scattering
spectra, with, except for region 1, a superimposed broad emission bump
in the red, which, at first sight, looks like
the ERE observed in galactic objects.

\begin{figure}[h]
\epsfxsize=9.0cm \epsfysize=16.0cm
\epsfbox{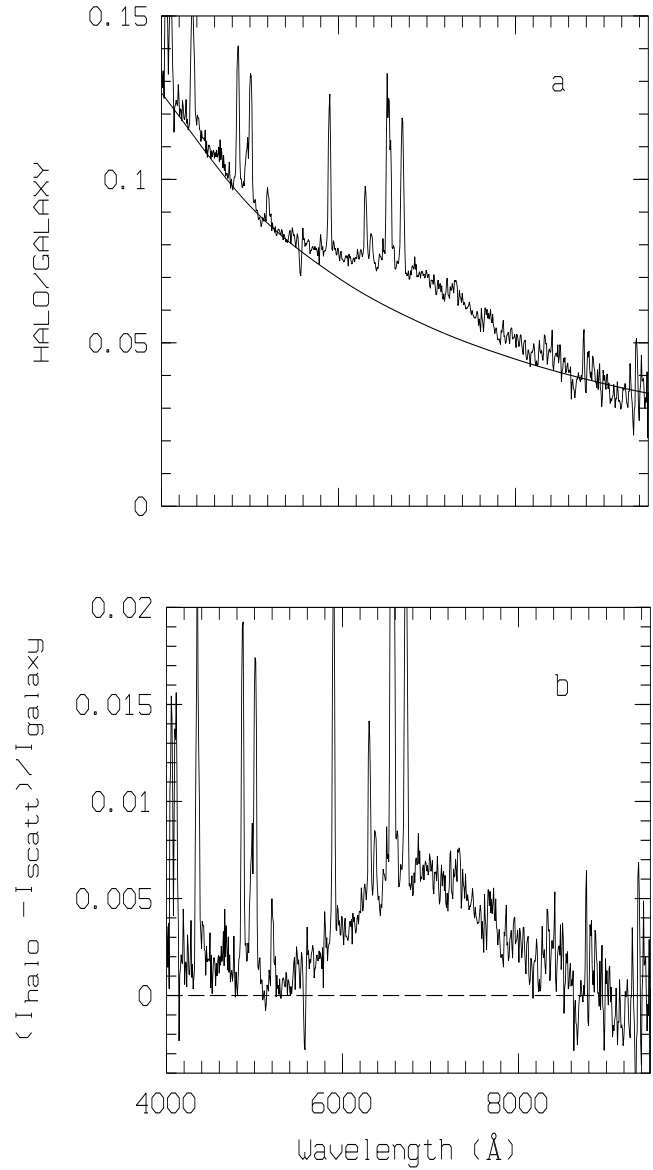}
\caption[]{   {\bf a}: Same figure as in Fig. \ref{divis}b. The spectrum
is well fitted in its bluer and redder parts by
a scattering continuum spectrum calculated for homogeneous HAC particles,
using complex indexes of refraction obtained by Khare et al. (1987)
for HACs synthesized in experimental conditions similar to those of
Furton \& Witt (1993).
{\bf b}: The ERE obtained by subtracting the calculated scattering
 continuum from the observed  spectrum}
\label{result}
\end{figure}

To isolate and characterize this bump, it is necessary to subtract the
scattering continuum. This could be done by calculating a polynomial fit
to the nebular spectrum divided by the galaxy, following the same method
as that of Witt \& Boroson (1990). We have prefered to fit a theoretical
scattering spectrum calculated for a real material: HAC grains
were chosen because they are a possible carrier of the ERE.
  Figure \ref{result}a shows one of the models best
fitting  the observed scattered component in the blue
($\lambda \la 5000$\AA) and in the red ($\lambda \ga 9000$\AA), two
ranges where no fluorescence is supposed to occur.

After subtraction of the fitted scattering component,
 a broad red bump appears (Fig.\ref{result}b). It is
centered at about 6900 \AA~ and is approximately 1500 \AA ~wide (FWHM).
This band thus appears to be markedly wider than the ERE observed in
galactic objects that generally does not exceed 1000 \AA ~in width.
Its central wavelength is about the same as that observed in the
reflection nebula NGC 2327 and in the planetary nebula NGC 7027,
which means that the peak wavelength is on the high side of the range
observed in these two categories of objects
 (Witt \& Boroson 1990, Furton \& Witt 1992). Comparison with galactic
H{\sc ii} regions (Perrin \& Sivan 1992, Sivan \& Perrin 1993)
shows that the ERE in M82 is bluer and, as in the previous cases, wider.

The ratio of ERE intensity  to  scattered light intensity,
integrated from 5500 to 8500 \AA~ is found to be
the same for regions 2, 3 and 4 and is about $0.150 \pm 0.005$.
This value can be compared to those of Witt \& Boroson (1990): it lies
in the middle of the range observed for
galactic reflection nebulae (if we except the Red Rectangle).

It should be noted that, of course, the fit to the observed continuum
obtained in Fig.  \ref{result}a, {\em does not rule out} other possiblities
for the nature of the scattering material  in the halo of M82.
It is shown simply that a population of HAC grains might provide
a possible common explanation for both luminescence and scattering,
an interesting result that could not have been obtained by a
simple analytical fit of the observed spectrum.
We have also considered scattering by pure amorphous carbon grains, using the
complex refractive indexes of Savvides (1986). No model was found to fit the
observations: this is consistent with the above result,  since
pure amorphous carbon grains  do not fluoresce in the red when irradiated
by ultraviolet photons (Furton \& Witt 1993).
\clearpage
\newpage
Silicates have been identified in M82 (Gillet et al. 1975).
Scattering spectra calculated using complex refractive indexes of
extra-terrestrial silicates (Egan \& Hilgeman 1979)
provide a good fit to the observations.
But the luminescence spectra of extra-terrestrial silicates is
markedly different from the observed ERE
(Derham \& Geake 1964, Derham et al. 1964).
Should organic ices and/or poly-HCN give rise to a luminescence band
lookliking the ERE, silicate grains coated with these materials
(Greenberg 1986) would explain completely our observations:
we have verified that their scattering spectra also provide a good
fit to the observed ones.
Relevant laboratory studies  on fluorescence from these materials induced
by ultraviolet light would prove to be very useful.
\verb+    +\\\\\\\\\\\\\\\\\\\\\\\\\\\\\\\\\\\\\\\\\\\\\\\\\\\\\\\\\\\\\\\\
\\\\\\\\\\\\\\\\\\\\\\\\\\\\\\\\
\section{Conclusion}
The observations reported in this Letter reveal for the first time the visible
continuum spectrum of some regions in the dusty halo of the galaxy M82. In
agreement with previous studies, it
is interpreted in terms of scattering by
grains of the light from the galaxy disk. Superimposed over the scattering
component, a broad emission band is found in the red, similar to, although
wider than the ERE observed in galactic reflection nebulae.

This is the first detection of ERE outside of our Galaxy.
\clearpage
\newpage
\it Acknowledgement. \rm We wish to thank E. Campinchi for her help regarding
the computations, G. Leydet for night assistance and Ph. Moreau for
photographic work.

\end{document}